\documentclass[conference]{IEEEtran}
\IEEEoverridecommandlockouts
\usepackage{cite}
\usepackage{amsmath,amssymb,amsfonts}
\usepackage{algorithmic}
\usepackage{tikz}
\usepackage{textcomp}
\usepackage{academicons}
\usepackage{colortbl}
\usepackage{hyperref}
\usepackage{pgf-pie}
\usetikzlibrary{trees}
\usepackage{forest}
\usepackage{graphicx}
\usepackage{pgfplots}
\usepackage{textcomp}
\usepackage{xcolor}
\usepackage{acronym}
\usepackage{multirow}
\usepackage{enumitem}
\usepackage{longtable}
\usepackage{threeparttable}
\usepackage{mathabx}
\usepackage{csquotes} 
\usepackage{tabularx,ragged2e,booktabs,caption}
\newcommand{\orcid}[1]{\href{https://orcid.org/#1}{\includegraphics[width=1em]{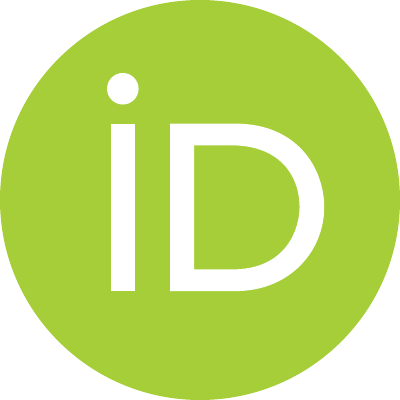}}}
\newcolumntype{C}{>{\centering\arraybackslash}X} 
\def\BibTeX{{\rm B\kern-.05em{\sc i\kern-.025em b}\kern-.08em
    T\kern-.1667em\lower.7ex\hbox{E}\kern-.125emX}}
\begin{document}

\title{Simulation-based application of Safety of The Intended Functionality to Mitigate Foreseeable Misuse in Automated Driving Systems\\

}

\author{\IEEEauthorblockN{1\textsuperscript{st} Milin Patel \orcid{0000-0002-8357-6018}}
\IEEEauthorblockA{\textit{Institute for Advanced Driver Assistance Systems and Connected Mobility} \\
Benningen, Germany\\ 
milin.patel@hs-kempten.de}
\and
\IEEEauthorblockN{2\textsuperscript{nd} Rolf Jung \orcid{0000-0002-0366-4844}}
\IEEEauthorblockA{\textit{Kempten University of Applied Sciences} \\
Kempten, Germany \\
rolf.jung@hs-kempten.de}
}

\maketitle

\begin{abstract}
The development of Automated Driving Systems (ADS) has the potential to revolutionize the transportation industry, but it also presents significant safety challenges. One of the key challenges is ensuring that the ADS is safe in the event of Foreseeable Misuse (FM) by the human driver. To address this challenge, a case study on simulation-based testing to mitigate FM by the driver using the driving simulator is presented. FM by the human driver refers to potential driving scenarios where the driver misinterprets the intended functionality of ADS, leading to hazardous behavior. Safety of the Intended Functionality (SOTIF) focuses on ensuring the absence of unreasonable risk resulting from hazardous behaviors related to functional insufficiencies caused by FM and performance limitations of sensors and machine learning-based algorithms for ADS. The simulation-based application of SOTIF to mitigate FM in ADS entails determining potential misuse scenarios, conducting simulation-based testing, and evaluating the effectiveness of measures dedicated to preventing or mitigating FM. The major contribution includes defining (i) test requirements for performing simulation-based testing of a potential misuse scenario, (ii) evaluation criteria in accordance with SOTIF requirements for implementing measures dedicated to preventing or mitigating FM, and (iii) approach to evaluate the effectiveness of the measures dedicated to preventing or mitigating FM. In conclusion, an exemplary case study incorporating driver-vehicle interface and driver interactions with ADS forming the basis for understanding the factors and causes contributing to FM is investigated. Furthermore, the test procedure for evaluating the effectiveness of the measures dedicated to preventing or mitigating FM by the driver is developed in this work.
\end{abstract}

\begin{IEEEkeywords}
ADS, FM, Simulation-based testing, SOTIF
\end{IEEEkeywords}

\section{Introduction}
\label{chap:1}
Automated Driving System (ADS) is a complex system that is designed to perform some or all of the driving tasks that are traditionally performed by a human driver \cite{Taxonomy_J3016_202104}. This includes acceleration, braking, steering, and navigation with the goal of improving road safety and reducing driver workload. Safety is crucial in ADS due to its reliance on perception sensors and complex algorithms for situational awareness. Malfunctions or failures in these systems can have hazardous consequences for vehicle occupants and other road users.

Foreseeable Misuse (FM) refers to the potential for human drivers to intentionally or unintentionally misuse ADS, leading to unsafe situations. This can occur when drivers do not fully understand the capabilities and limitations of the ADS, or when they engage in behaviors that are not consistent with the intended use of the system. FM can also occur when drivers fail to take over control of the vehicle when required, such as in situations where the system is unable to operate safely, or when the driver is required to take over control due to a system malfunction. \cite{ISO.21448}

The safety challenges associated with FM are significant because they can lead to serious accidents and injuries. To address these challenges, it is important to develop and implement effective mitigation measures that can prevent or reduce the risk of FM. This requires a comprehensive evaluation of the ADS that considers a wide range of driving scenarios and human factors, including the driver's understanding of the system's capabilities and limitations, their responsibilities, and their ability to comprehend and respond to warnings and alerts

To illustrate the significance of addressing FM, consider the fatal accident involving a Tesla Model S in 2016. The driver was using the Autopilot system, which is intended to assist with steering, braking, and acceleration. However, the driver was not paying attention to the road and did not take over control of the vehicle when required. As a result, the autopilot system failed to detect a truck that was crossing the road, and the vehicle collided with the truck, resulting in the driver's death. \cite{NTSB.}
                                                         
ISO 21448 \cite{ISO.21448} is the standard for Safety of the Intended Functionality (SOTIF) that provides guidance to identify and analyze potential hazards and risks associated with the intended functionality of an ADS that may arise due to foreseeable misuse by human drivers. In the following, the term ‘system’ is used in place of ADS.

Testing for FM is a challenging task due to the system's complexity and the vast range of potential misuse scenarios. Anticipating all potential misuse scenarios poses challenges in designing and testing the system to effectively prevent or mitigate such occurrences. Additionally, testing for FM can be time-consuming and expensive, requiring extensive testing and evaluation to ensure that the system meets intended safety requirements.

Simulation-based testing overcomes the above-mentioned challenges by creating a controlled environment to systematically evaluate potential misuse scenarios, replicating challenging or hazardous real-world scenarios. Simulation-based testing offers a safer and more efficient means of comprehensively evaluating a system's responses, robustness, and ability to effectively prevent or mitigate misuse effects.

Patel et al. \cite{10.1007/978-981-99-3043-2_15} proposed a simulation-based approach for testing FM by the driver in highly automated driving systems and discussed the importance of managing driver-system interactions and the implications of driver-vehicle-interface design on these interactions. However, the proposed approach in the paper \cite{10.1007/978-981-99-3043-2_15} does not focus on the mitigation of FM.

Mitigation of FM refers to the process of identifying potential misuse of the intended functionality of the system and taking measures to reduce the associated risks to an acceptable level. 

The major contribution of the presented work lies in defining the simulation-based test procedure for evaluating the effectiveness of the measures dedicated to preventing or mitigating FM by the driver. In this context, the main contributions of this work are to:
\begin{enumerate}[label=\roman*.]

\item Define the test requirements: The test requirements must specify the simulation environment, including the vehicle model, the sensor models, and the simulation software. The test requirements must be designed to ensure that the simulation accurately represents the real-world scenario and that the system's response to the scenario can be evaluated. 
	
\item Formulate evaluation criteria: The evaluation criteria must be designed to ensure that the measures are effective in preventing or mitigating FM and that they do not introduce new hazards or adversely affect the vehicle's intended functionality. 
	
\item Propose an approach to evaluate the effectiveness of measures dedicated to preventing or mitigating FM: The approach must include testing the system's response to potential misuse scenarios, to ensure that the system can detect and respond appropriately to these scenarios. 
\end{enumerate}
\subsection{Structure of the paper}
The subsequent chapters of this paper are organized as follows: Chapter \ref{chap:2} presents an extensive background, briefly explaining misuse types in ADS, factors and causes of misuse, accompanied by a case study on simulation-based testing of foreseeable misuse. In Chapter \ref{chap:3}, the concept for simulation-based application is proposed, detailing the workstation setup and defining essential test requirements. Chapter \ref{chap:4} delves into approach to evaluate the effectiveness of measures dedicated to prevent or mitigate FM, emphasizing conditional probability analysis, Foreseeable Misuse Evaluation Metric (FMEM), and the assessment of simulation results. Finally, Chapter \ref{chap:5} concludes by summarizing key findings and implications, and suggesting future research directions.

\section{Background}
\label{chap:2}
\subsection{Types of Misuse}
ISO 21448 \cite{ISO.21448} defines two types of misuse in the ADS: direct misuse and indirect misuse. Direct misuse involves conditions that can trigger hazardous behavior in the system, while indirect misuse refers to driver behavior that reduces controllability or increases the severity of an accident without directly causing hazardous behavior in the system.

Instances of direct misuse in ADSs include overconfidence in system performance, misunderstanding of the system's capabilities, lack of understanding regarding system functions, incorrect assumptions about driver interaction based on design specifications, and driver expectations that do not align with the system's capabilities. \cite{ISO.21448}

Indirect misuse instances involve driver fatigue leading to decreased ability to interact with or monitor automation features, distractions stemming from mobile devices or other passengers, reduced attentiveness due to prolonged use or monotonous driving conditions, and over-reliance on automated driving functions without maintaining situational awareness. \cite{ISO.21448}

\subsection{Factors and Causes of misuse}
Driver Recognition (DR), and Driver Judgment (DJ) are factors and causes of misuse that can contribute to direct or indirect misuse scenarios.

\subsubsection*{Driver Recognition (DR)}
Driver recognition refers to the process of perceiving and interpreting the driving environment, including the road, traffic, and surrounding objects. Failures in driver recognition may manifest when drivers overlook the limitations of ADS or misinterpret the driving environment.

For example, a driver may fail to recognize that the ADS is not capable of detecting certain objects like pedestrians or bicycles, and may rely on the system to avoid collisions. This can lead to unsafe situations if the ADS fails to detect these objects, and the driver does not take appropriate action. \cite{SOTIF.lanecentering.highwayschauffeursystem}

\subsubsection*{Driver Judgment (DJ)}
Driver judgment refers to the process of making decisions based on the driving environment and the capabilities of the ADS. FM related to driver judgment can occur when drivers make erroneous decisions based on incorrect assumptions about the ADS or the driving environment.

For example, a driver assuming that the ADS is capable of navigating through heavy rain or fog, even though the system is not designed for this purpose. The ADS may not be able to detect the road markings or other objects in such conditions, and the driver may not take appropriate action, leading to a collision. \cite{SOTIF.lanecentering.highwayschauffeursystem}

\subsection{Case Study on Simulation-based testing for Foreseeable Misuse by the driver}
Patel et al. \cite{10.1007/978-981-99-3043-2_15} proposed a strategy for implementing simulation-based testing of FM resulting from the system-initiated transition between the human driver and the ADS. The system-initiated transition is the process and period for transferring responsibility and driving control over some or all aspects of the driving tasks between the human driver and the system.

Simulation-based testing involves using a driving simulator to simulate a modeled misuse scenario in the virtual test environment and analyzing the results to determine whether the system meets the intended safety requirements. However, simulation can be limited by the underlying assumptions about environmental conditions, sensors, and the vehicle model.

Patel et al. \cite{10.1007/978-981-99-3043-2_15} acknowledged that the strategy presented is to demonstrate an approach for simulation-based testing of FM and is not intended to be a distinctive or optimal measure dedicated to mitigating FM. Also, the implementation of the strategy has not been evaluated in practice, and therefore, the effectiveness of the measures dedicated to preventing or mitigating FM has not been evaluated.
\section{Proposed concept for Simulation-based Application}
\label{chap:3}
\subsection{Determining SOTIF-related Misuse Scenario}
\label{det:scenario}
To derive a misuse scenario, various sources, including lessons learned, expert knowledge, and brainstorming, can be utilized. ISO 21448 (Annex B1) provides a systematic approach for deriving an SOTIF-related misuse scenario. \citen{ISO.21448}

The process for identifying a misuse scenario begins with understanding the intended functionality of the system. This entails comprehending the system's purpose, its designated user base, and the environment in which it will operate. Once the intended functionality is understood, the next step is to identify potential misuses of the system. This includes understanding how the system could be used in unintended ways and how these misuses could lead to hazards. \cite{inproceedings}

The potential misuses can be categorized based on their severity and likelihood. This helps prioritize the misuses that pose the greatest risk to safety. Based on the potential misuses identified, a misuse scenario should be developed. The misuse scenario should describe how the system could be misused, the potential consequences of the misuse, and the likelihood of the misuse occurring. \cite{inproceedings}

Lastly, the evaluation of the misuse scenario is crucial to determine its impact on the safety of the ADS. This includes assessing the likelihood of the misuse occurring, the severity of the consequences, and the effectiveness of any mitigation measures that could be implemented. It is important to note that misuse scenarios should not only consider deliberate violations but also human driver errors that could lead to the unintended use of the system.

The Table \ref{scenariotable} depicts a description of a misuse scenario derived from \cite{10.1007/978-981-99-3043-2_15}, in accordance with an example methodology outlined in ISO 21448 (Annex B1).

\begin{table}[!htbp]
	\renewcommand{\arraystretch}{1.25}
	\setlength{\tabcolsep}{2.5pt}
	   \begin{threeparttable}
	\caption{Description of SOTIF-related misuse scenario, adapted from \cite{ISO.21448}}
\label{scenariotable}
	\begin{tabular}{|p{1.125cm}|p{1.125cm}|p{1.125cm}|p{1.125cm}|p{1.125cm}|p{1.1cm}|p{1.10cm}|}
		\hline
	Potential SOTIF-related misuse scenario & Stake-holder                                                
			&  \multicolumn{2}{p{1cm}|}{Foreseeable Misuse}                             
			& 
			Driver-System Interactions
			&   Environ- \newline mental  Conditions
			&    Derived Hazardous Scenario
			
			\\ \cline{3-4} &  &  Factors &  Causes &  & &
			   
			 	\\ \hline
\multirow{4}{2cm}{Described \newline below \tnote{1}} & \multirow{4}{1cm}{Driver}                  
			&
			Recogni \newline tion 
			& 
		   False recognition
			& \multirow{4}{2cm}{Described \newline below \tnote{2}}
			&  \multirow{4}{2cm}{Described \newline below \tnote{3}}
			 & \multirow{4}{2cm}{Described \newline below \tnote{4}}
			 
			\\ \cline{3-4}
			& 	&  
			Judgement 
			& Misjudg \newline ment & & &		
\\ \hline
		\end{tabular}	
	
	\begin{tablenotes}
		\item[1] The Ego-Vehicle encounters a road with missing lane markings during automated driving on a two lane one-way highway and executing lane change maneuver from right to left lane. The camera sensor cannot estimate the location of the lane boundary due to a performance limitation of the camera sensor. Ego-Vehicle starts to leave the lane and driver is notified to take control of the driving tasks by means of Take-Over-Request (TOR).
		
		\item[2] \enquote{Delayed Take-over} and/or  \enquote{Take-Over and perform Over/Understeer}
		
		\item[3] \begin{itemize} 	\item Weather : clear  	\item Light Condition : daylight 	\item Traffic Condition : light traffic 	\item Roadway Surface and Features : missing lane markings 	\end{itemize}
		\item[4] Driver fails to take-over the control of the driving tasks, resulting in lane departure of Ego-Vehicle.
	
	\end{tablenotes}
\end{threeparttable}
\end{table}

\subsection{Workstation setup for Simulation-based Testing}
The workstation has been developed with an integrated driving simulator equipped with hardware tools, including the Logitech G29 steering wheel, pedals, and gearbox, integrated with a simulation tool, IPG CarMaker, to perform simulation-based testing of FM. The driving simulator is a static simulator that allows a human driver to engage in the simulation-based application.

\begin{figure}[h]
	\centering
	\includegraphics[width=\linewidth,height=0.35\textheight]{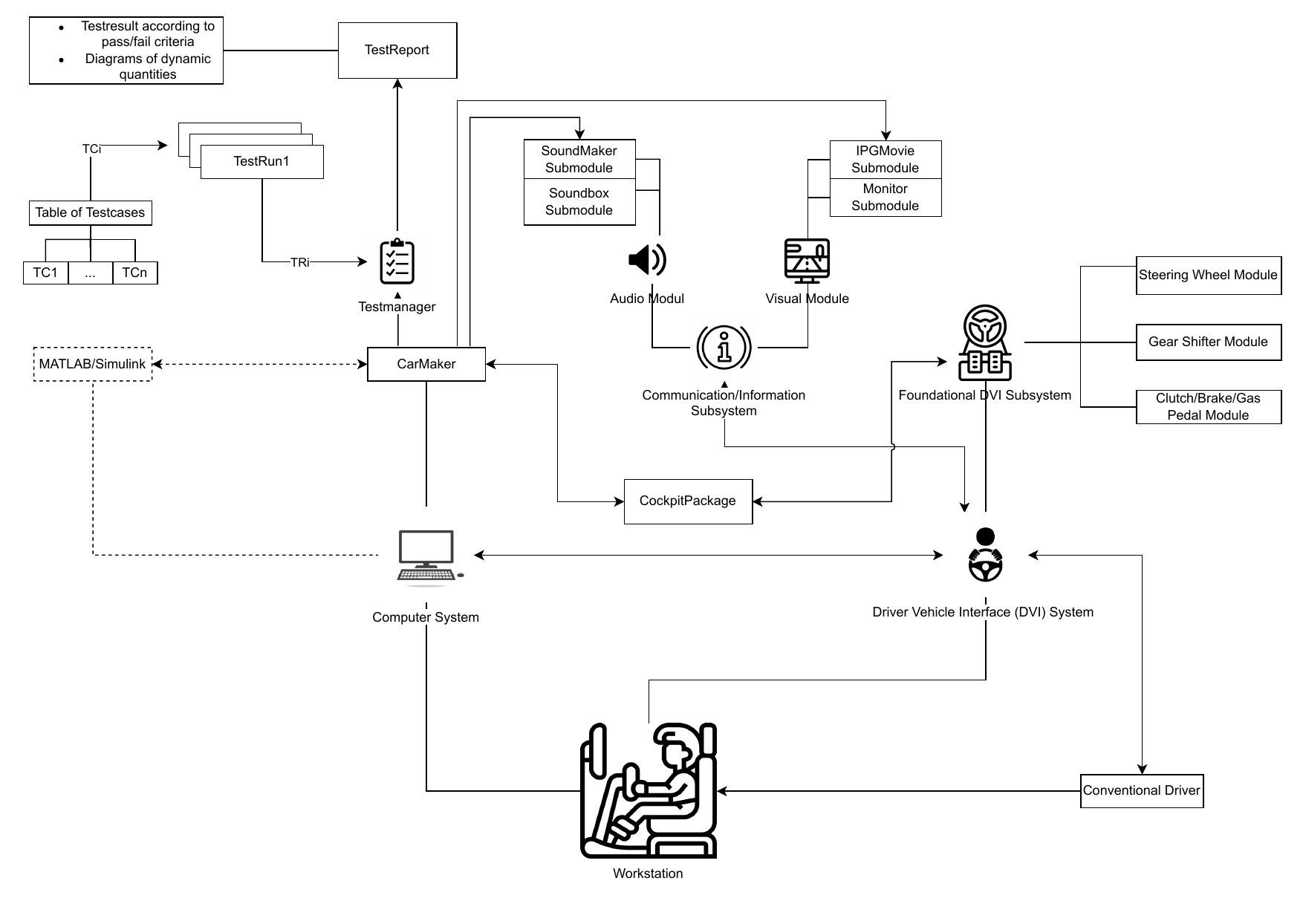}
	\caption{Representation of a workstation setup to perform simulation-based testing of foreseeable misuse by the driver}
	\label{fig:driving simulator}
\end{figure}

The block diagram in Figure \ref{fig:driving simulator} represents a workstation setup consisting of several components that collaborate to execute driving tasks.

The \enquote{Foundational DVI subsystem} is responsible for manual driving and allows the driver to interact with sensor module, controller module, environment module, and other vehicle modules.

\enquote{CarMaker} is a software tool for developing and testing virtual vehicle models, while \enquote{CockpitPackage} is an extension in CarMaker that facilitates the integration of hardware components including steering wheels, clutch/brake/gas pedals, and gear shifters into the simulation environment.

The \enquote{Communication/Information} subsystem provides visualization of the simulation and includes modules like IPGMovie for real-time 3D animation of the virtual driving simulation and Sound-Maker software for 3D audio. It also encompasses the Visual Module, which contains the Instruments Panel sub-module for providing information to the driver about the current status of the driving mode and warnings in case of upcoming events.

TestManager is a software tool employed in CarMaker for simulating Test Cases (TC). It permits the creation, execution, and management of TC.
The test cases are a set of predefined procedures that are used to verify the functionality of the system under test. The TestManager sequentially invokes the TestRuns to execute a Test-case Series (TCS). After each TCS, a TestReport is generated, displaying all executed TC and their outcomes. The TestManager also allows the addition of pass/fail criteria to determine TCS success.

\subsection{Defining test requirements for performing simulation-based testing}
\label{def_requirements}
The simulation requirements become relevant at different phases of simulation-based testing. Therefore, test requirements are formulated in form of a checklist, encompassing a series of questions. The checklist is presented in form of question cards.

Table \ref{tab:requirementtabeltemplate} illustrates the template of the presented question cards. In the top-right corner of the question card, the question number  is denoted by \textbf{[1, 2, ..., n]}, followed by the primary question. On the bottom-left, possible responses to the question are provided. An explanation for the respective question is provided in the bottom-right corner of the table.

\begin{table}[!htbp]
	\renewcommand{\arraystretch}{1.25}
	\setlength{\tabcolsep}{3pt}
	\caption{Template of the checklist}
	\label{tab:requirementtabeltemplate}
	\begin{tabular}{|p{2.8cm}|p{5.5cm}|}
		\hline
		
		Question Number &  \textbf{[1,2,…,n]} \\
		\hline
\multicolumn{2}{|c|}{Main Question}	  \\ 
		\hline
		Possible Response  & Explanation\\	\hline
	
	\end{tabular}
\end{table}	
In the following, a set of questions (No.\textbf{1} – No.\textbf{10}) representing the requirements for  the simulation-based testing are provided in Table \ref{tab:requirementtable} below.
\begin{table}[!htbp]
	\renewcommand{\arraystretch}{1.25}
	\setlength{\tabcolsep}{3pt}
	\caption{Question Cards}
	\label{tab:requirementtable}
	\begin{tabular}{|p{2.8cm}|p{5.5cm}|}
		\hline
		Question Number &  \textbf{1} \\
		\hline
		\multicolumn{2}{|p{8.3cm}|}{Does the system performs automated driving of the ego-vehicle by providing longitudinal and lateral control in the modeled scenario?}	  \\ 
		\hline
		Yes / No
		& The simulation starts initially in automated driving mode. It is intended to provide longitudinal control of the Ego-Vehicle on a right lane in a one way two-lane highway environment. All automated driving function should be active and work.
		\\	\hline
		
		Question Number &  \textbf{2} \\
		\hline
		\multicolumn{2}{|p{8.3cm}|}{Does the Ego-Vehicle encounters a road with unclear lane markings during the initialization of the lane change from right lane to left lane?}	  \\ 
		\hline
		
		Yes / No
		& The Ego-Vehicle encounters a part of the road with unclear lane markings while executing lane change maneuver from right to left lane.
		\\	\hline
		
		Question Number &  \textbf{3} \\
		\hline
		\multicolumn{2}{|p{8.3cm}|}{Does the ADS send a warning to the driver when the Ego-Vehicle encounters the specified road conditions in Question 3?}	  \\ 
		\hline
		
		Yes / No
		& Visually and auditory warning is sent out at 6.04 seconds of the simulation time and the lane departure warning is activated because unclear lane markings are detected.
		\\	\hline
		
		Question Number &  \textbf{4} \\
		\hline
		\multicolumn{2}{|p{8.3cm}|}{If the driver not respond to the warnings, does the system notify the driver by issuing imminent Take-Over-Request (TOR)?}	  \\ 
		\hline
		
		Yes / No
		& If the driver not response to the warning then the ADS will request the driver to Take-Over (TO). The TOR is sent at 7.96 seconds of the simulation time.
		\\	\hline
		
		Question Number &  \textbf{5} \\
		\hline
		\multicolumn{2}{|p{8.3cm}|}{Does the driver response to the TOR?}	  \\ 
		\hline
		
		Yes / No
		& The driver can respond to the TOR and does the TO of driving task by means of driver-vehicle-interface (Logitech G29 Steering Wheel buttons). The system is expected to remain operational in automated driving mode until the driver is able to regain control of the driving task. 
		
		\\	\hline
		
		Question Number &  \textbf{6} \\
		\hline
		\multicolumn{2}{|p{8.3cm}|}{Does the system transition into automated driving with reduced functionality mode and performs minimal risk maneuver?}	  \\ 
		\hline
		
		Yes / No
		& If the driver does not TO the driving tasks in the event of TOR, the system will transition to the automated driving with reduced functionality. Subsequently, a minimal risk maneuver is performed by the system to keep the Ego-Vehicle in	its lane and to automatically stop the Ego-Vehicle on the side of the road in a safe manner \cite{safetyfirst}.
		\\	\hline

		Question Number &  \textbf{7} \\
		\hline
		\multicolumn{2}{|p{8.3cm}|}{Does the driver TO of the driving task in the specified time?}	  \\ 
		\hline
		
		Yes / No
		& The driver does TO in time, if the TO time is less than 1.77 seconds \cite{time.takeover}. The TO-time is the difference between TOR and TO. TO after 1.77 seconds are considered as delayed TO (i.e., considered as FM).
		\\	\hline
		
		Question Number &  \textbf{8} \\
		\hline
		\multicolumn{2}{|p{8.3cm}|}{Does the adjusted Steering wheel Angle (SWA) by the driver after TO lead to Oversteer or Understeer (i.e, considered as FM) ?}	  \\ 
		\hline
	
		Yes / No
		& 	Ideal SWA could be defined as the SWA that will centre the Ego-Vehicle in the middle of the current lane (left lane). \newline
		•	Over-steer if the adjusted SWA is greater than ideal SWA. \newline
		•	Understeer if the adjusted SWA is smaller than ideal SWA. 
		\\	\hline
			\end{tabular}
	\end{table}

\begin{table}[!htbp]
\renewcommand{\arraystretch}{1.25}
\setlength{\tabcolsep}{3pt}
\label{tab:requirementtable-continued}
\begin{tabular}{|p{2.8cm}|p{5.5cm}|}

\multicolumn{2}{p{8.3cm}}{\small\textit{Table \thetable, continued from previous page}} \\
\hline
		Question Number &  \textbf{9} \\
		\hline
		\multicolumn{2}{|p{8.3cm}|}{Does the FM by the driver lead to Hazard?}	  \\ 
		\hline
		
		Yes / No
		&  Hazard if the Ego-Vehicle departs lane. \newline
		a.	Lane departure towards the east from left lane 	or \newline
		b.	Lane departure towards the west from left lane
		\\	\hline
		
		Question Number &  \textbf{10} \\
		\hline
		\multicolumn{2}{|p{8.3cm}|}{Does the driver able to handle the driving situation (after TO by driver), measured as controllability?}	  \\ 
		\hline

		Yes / No
		&  Likelihood that the driver can cope with driving situations including the system limits and system failures is defined as \enquote{controllability} \cite{Knapp.2009}. \newline
		• Controllability is provided if driver does TO successfully without leading to hazard. \newline
		• Controllability is not provided if driver TO unsuccessfully leading to hazard.
		\\	\hline
		
	\end{tabular}
\end{table}	

\newpage
\section{Approach to evaluate the effectiveness of measures dedicated to preventing or mitigating Foreseeable Misuse}

\label{chap:4}
\subsection{Conditional Probability Analysis}
\label{CPA}
Conditional Probability Analysis (CPA) is chosen as the primary approach to establish a systematic and quantifiable approach to understanding relationships between factors and causes of FM within the ADS. In particular, the focus is on two key elements: False Recognition (FR) and Misjudgment (MJ).

The choice of employing the CPA approach is informed by the methodology outlined by Mkrtchyan et al. \cite{MKRTCHYAN201693}, which demonstrates the applicability of this approach in evaluating safety-critical systems.

The below provided analysis is based on considering the SOTIF-related misuse scenario described in Table \ref{scenariotable} (\ref{det:scenario}), and in accordance with defined requirements in Table \ref{tab:requirementtable} (\ref{def_requirements}).

\subsubsection*{\textbf{Misjudgment (MJ)}}
It relates to situations where the driver makes an erroneous decision during the Take-Over (TO) process, potentially resulting in under-steering or over-steering, and potential for a lane departure (i.e., hazard).

The probability of Misjudgment is assessed under two conditions: 
\begin{enumerate}
	\item when there is no delayed take-over ($TO \leq 1.77$ seconds) and a Hazard (H) is present, mathematically expressed as: 	
\begin{equation}
	P(MJ | TO \leq 1.77\,s, H) = \frac{P(TO \leq 1.77\,s \cap H)}{P(MJ \cap TO \leq 1.77\,s \cap H)}
\end{equation}

	\item  when there is a delayed take-over ($TO > 1.77$ seconds) in the presence of a Hazard (H), mathematically expressed as: 
\begin{equation}
	P(MJ | TO > 1.77\,s, H) = \frac{P(TO > 1.77\,s \cap H)}{P(MJ \cap TO > 1.77\,s \cap H)}
\end{equation}
\end{enumerate}

\subsubsection*{\textbf{False Recognition (FR)}}
It occurs when the driver fails to promptly recognize the necessity of take-over control. This can result in delayed takeover, leading to a delayed take-over and the potential for a lane departure (i.e., hazard).

The probability of False Recognition is evaluated under the condition of a delayed take-over ($TO > 1.77$ seconds) when a Hazard (H) is present, mathematically expressed as: 
\begin{equation}
	P(FR | TO > 1.77\,s, H) = \frac{P(TO > 1.77\,s \cap H)}{P(FR \cap TO > 1.77\,s \cap H)}
\end{equation}

\begin{figure}[t]
	\centering
	\caption{Probability Tree Diagram for Misjudgment (MJ) and False Recognition (FR)}
	\label{Treediagram}
	\resizebox{\columnwidth}{!}{%
		\begin{forest}
			for tree={
				draw,
				align=center,
				parent anchor=south,
				child anchor=north,
				edge={->},
				l sep+=10mm,
				s sep+=20mm,
				fit=band,
			}
			[MJ
			[TO $\leq$ 1.77s
			[H
			[FR $|$ TO $\leq$ 1.77s \& H]
			]
			[No H
			[FR $|$ TO $\leq$ 1.77s \& No H]
			]
			]
			[TO $>$  1.77s
			[H
			[TO $>$  1.77s \& H]
			]
			[No H
			[TO $>$ 1.77s \& No H]
			]
			]
			]
		\end{forest}
	}
	
	\vspace{20pt} 
	
	\begin{minipage}{\columnwidth}
		\footnotesize
		\begin{itemize}
			\item The top level represents the conditional probability of Misjudgment (MJ).
			\item	The second level represents the two scenarios based on Takeover  (TO) time being less than or greater than 1.77 seconds.
			\item	The third level represents the presence or absence of Hazard (H) in each scenario.
			\item	The last level represents the conditional probability of False Recognition (FR) based on the Takeover (TO) time and Hazard (H) conditions.
		\end{itemize}
	\end{minipage}
	
\end{figure}
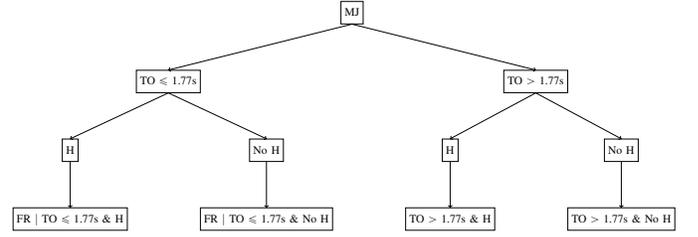

As a visual aid, Figure \ref{Treediagram} illustrates a probability tree diagram depicting the relationships between Misjudgment (MJ) and False Recognition (FR). This diagram provides a structured representation of conditional probabilities, enhancing the understanding of the interactions among various factors and causes of FM.

\subsection{Foreseeable Misuse Evaluation Metric (FMEM)}
The Foreseeable Misuse Evaluation Metric (FMEM)  systematically categorizes instances into four distinct groups: True Positives, False Positives, True Negatives, and False Negatives. 

The FMEM metric for assessing the effectiveness of ADS in identifying and mitigating instances of FM is defined as follows, adapted from \cite{Ting.2010}:

\begin{equation}
	\label{FMEMEQ}
	FMEM = \frac{TP + TN}{TP + TN + FP + FN}
\end{equation}

\subsubsection*{\textbf{Metric Definitions}}
\label{sec:definitions}
The following definitions apply to each FMEM group:

\begin{itemize}
	\item True Positive (TP): Instances where the ADS accurately identifies and mitigates FM in situations where a hazard is present and the takeover time exceeds a predefined threshold.
	
	\item False Positive (FP): Instances where the ADS erroneously detects FM in the absence of a hazard or when the takeover time is below or equal to the established threshold.
	
	\item True Negative (TN): Instances where the ADS correctly concludes the absence of FM, either when there is no hazard present or when the takeover time surpasses the threshold.
	
	\item False Negative (FN): Instances where the ADS fails to detect FM despite the presence of a hazard and the takeover time exceeding the threshold.
\end{itemize}

The equation \ref{FMEMEQ} represents the accuracy of the ADS, defined as the proportion of correctly identified instances (both True Positives and True Negatives) out of all evaluated instances. Here, TP and TN are instances where the system’s response aligns with the actual situation (whether identifying a hazard or its absence correctly), while FP and FN are instances where the system’s response does not align with actual situation.

\subsection{Evaluation of simulation-based application}
An exemplary Table \ref{tab:tcsexample} of Test Case Series (TCS) is presented based on results of simulation-based application of FM. Because of the high amount of TC, all simulation results of TC are not entailed. Only a exemplary Table \ref{tab:tcsexample} is provided to demonstrate the results.

TestManager is a tool employed in software CarMaker for simulating Test Cases (TC). TestManager allows the creation, execution, and management of TC. The TestManager sequentially invokes the TC to execute a Test-case Series (TCS).

\begin{table}[h]
	\centering
	\caption{Exemplary Test Case Series (TCS)}
	\label{tab:tcsexample}
		\renewcommand{\arraystretch}{1.25}
	\setlength{\tabcolsep}{1pt}
	\begin{tabular}{|c|c|c|c|c|c|c|c|c|c|}
	\hline
		TC & TO & TO\_t2 [s] & delta\_T2 [s] & DelTO & SWA [deg] & H & H\_t3 [s] & delta\_T3 [s] \\ 
		\hline
		1 & 1 & 10.2300 & 2.2700 & 1 & 12.5144 & \cellcolor{green} 0 & 0.0000 & 0.0000 \\ \hline
		2 & 1 & 10.7300 & 2.7700 & 1 & 3.2086 & \cellcolor{green} 0 & 0.0000 & 0.0000 \\ \hline
		3 & 1 & 11.0800 & 3.1200 & 1 & 15.2058 & \cellcolor{red} 1 & 11.1000 & 0.0200 \\ \hline
		4 & 1 & 9.1200 & 1.1600 & 0 & 32.1657 & \cellcolor{red} 1 & 10.4000 & 1.2800 \\ \hline
		5 & 1 & 11.3500 & 3.3900 & 1 & 10.5064 & \cellcolor{red} 1 & 11.1000 & 0.4600 \\
		\hline
		... & ... & ... & ... & ...& ... & ... & ... & ... \\
		\hline
		TC$_n$ & ... & ... & ... & ...& ... & ... & ... & ...\\
		\hline
	\end{tabular}
\end{table}

The Test Cases (TC) encompass various parameters, including Takeover (TO), Delayed TO (DelTO), and Hazard (H), which are logic values, either 0 or 1. The explanations for each parameter are presented below.

\subsubsection{Takeover (TO)}
The parameter \enquote{TO} signifies whether the driver take-over control of the driving task in a given TC. A value of 1 indicates a take-over of the driving task by the driver.

\subsubsection{Delayed Takeover (DelTO)}
The parameter \enquote{DelTO} is set to 1 when the driver take-over control after or/ more than 1.77 seconds. To calculate DelTO, \enquote{delta\_T2} is used, representing the time difference between the Takeover Request (TOR) time (fix value = 7.96 seconds \cite{10.1007/978-981-99-3043-2_15}) and the actual Takeover time (TO\_t2). For instance, in TC 1, \enquote{DelTO} is 1 because the driver does take-over after 2.27 seconds.

\subsubsection{Hazard (H)}
The parameter \enquote{H} indicates the lane departure of the vehicle(i.e., hazard) during a TC. It equals 1 if a hazard occurs.  For instance, In TC 3, TC4 and TC5, \enquote{H} is 1 that indicates a lane departure.

\subsubsection{Hazard Time (H\_t3) and Delta\_T3}
The parameter \enquote{H\_t3} represents the time of a hazard in a TC, and \enquote{Delta\_T3} is the time difference between \enquote{H\_t3} and \enquote{TO\_t2}. In TC 3, \enquote{H\_t3} is 11.1 seconds, and \enquote{Delta\_T3} is 0.02 seconds.

\subsubsection{Steering Wheel Angle (SWA)}
The parameter \enquote{SWA} reflects the driver's steering input at take-over. It is calculated as the difference between SWA at \enquote{TO\_t2} and the maximum steering input immediately after take-over. 

\subsection{Assessment of Simulation Results}

As mentioned in requirement Table \ref{tab:requirementtable}, question 10, likelihood that the driver can cope with driving situations including the system limits and system failures is defined as \enquote{controllability}. 

Within the simulation-based application, a total of 50 Test Cases (TC) were conducted. Among these, 22 TC (44\%) exhibit controllability, signifying that the driver could regain control of the vehicle after a take-over. Conversely, 28 TC (56\%) revealed a lack of controllability, resulting in hazard.

The Table \ref{tab:before-modification} presents probability analysis results based on simulation results. It outlines the likelihood of different events related to driver controllability. The outcomes reflect the chances of a specific scenario, including Misjudgment (MJ) and False Recognition (FR) under different driving conditions:

\begin{table}[!htbp]
	\renewcommand{\arraystretch}{1.25}
	\setlength{\tabcolsep}{3pt}
	\caption{Probability Analysis Results Based on Simulation Data}
	\label{tab:before-modification}

	\begin{tabular}{|p{3.5cm}|p{2.3cm}|p{2.5cm}|}
		\hline
	  Probability Condition & Number of Test Cases & Percentage\\
		\hline
		P(TO $\leq$ 1.77s $\cap$ H) & 10 & 20\% \\ \hline
		P(MJ $\cap$ TO $\leq$ 1.77 s $\cap$ H) & 6 & 12\% \\ \hline
		P(TO $>$ 1.77 s $\cap$ H) & 2 & 4\% \\ \hline
		P(MJ $\cap$ TO $>$ 1.77 s $\cap$ H) & 8  & 16\% \\ \hline
		P(FR $\cap$ TO $>$ 1.77 s $\cap$ H) & 2 & 4\% \\ \hline
	\end{tabular}
\end{table}

The conditional probabilities, as outlined in Table \ref{tab:before-modification}, can be calculated for Chapter \ref{CPA}.

\subsubsection*{Misjudgment (MJ) Probability}

\begin{enumerate}
	\item When there is no delayed take-over ($TO \leq 1.77$ seconds) and a Hazard (H) is present:
	\begin{equation}
		P(MJ | TO \leq 1.77\,s, H) = \frac{10}{6} = 1.67.
	\end{equation}
This signifies that a driver is 1.67 times more likely to misjudge a situation when there is no delayed take-over ($TO \leq 1.77$ seconds) and a Hazard (H) is present.
	
	\item When there is a delayed take-over ($TO > 1.77$ seconds) in the presence of a Hazard (H):
	\begin{equation}
		P(MJ | TO > 1.77\,s, H) = \frac{2}{8} = 0.25.
	\end{equation}
In this case, the likelihood of misjudgment is significantly reduced when there is a delayed take-over ($TO > 1.77$ seconds) in the presence of a Hazard (H).
	
\end{enumerate}

\subsubsection*{False Recognition (FR) Probability}
The probability of False Recognition is evaluated under the condition of a delayed take-over ($TO > 1.77$ seconds) when a Hazard (H) is present:

\begin{equation}
	P(FR | TO > 1.77\,s, H) = \frac{2}{2} = 1.
\end{equation}
This implies that the probability of false recognition remains at 100\% when there is a delayed take-over ($TO > 1.77$ seconds) and a Hazard (H) is present, indicating a high likelihood of recognizing a false situation.

It is important to note that these calculated probabilities are specific to particular conditions and a scenario with a limited number of test cases used in the simulation. These values are derived based on a specific threshold of 1.77 seconds for take-over time, and results may vary with different thresholds. Additionally, these results do not account for other potential factors that could influence driver behavior.

\section{Conclusion and Future Work}
\label{chap:5}
The paper presents a detailed strategy for evaluating the effectiveness of measures aimed at preventing or mitigating Foreseeable Misuse (FM) within Automated Driving Systems (ADS). The predominant method used for evaluation is Conditional Probability Analysis (CPA), concentrating on principal factors and causes for FM, such as False Recognition (FR) and Misjudgment (MJ). This application of CPA is specifically applied to a misuse scenario related to the Safety of the Intended Functionality (SOTIF).

The analysis yields insightful findings regarding the conditional probabilities of Misjudgment and False Recognition occurring under various driving conditions. It is observed that the probability of Misjudgment significantly reduces when Take-Over (TO) is delayed, particularly in the presence of a Hazard (H). However, the probability of False Recognition remains substantially high when TO is delayed, leading to a Hazard (H).

Although these findings provide meaningful insights into FM evaluation within ADS, it is crucial to recognize that the calculated probabilities are particular to a specific misuse scenario and the limited number of test cases used in the simulation. These figures are a fundamental basis for further detailed exploration and adaptation to actual scenarios, considering variations in threshold values for TO and other influential factors that affect driver behavior.

For future research, there are several promising directions. Initially, it is imperative to gather and analyze real-world data to verify the simulation-based results. Validating these findings in real driving conditions is essential for an accurate understanding of Foreseeable Misuse (FM).

Furthermore, it is vital to conduct sensitivity analyses on the threshold value for Take-Over (TO) time. Investigating how changes in the TO threshold impact conditional probabilities will provide deeper insights. Additionally, a comprehensive study of driver behavior and cognitive processes during take-over events is necessary. Such research should investigate factors like driver fatigue, distractions, and experience, and their significant impact on the likelihood of FM.

\bibliographystyle{IEEEtran}
\bibliography{mybibfile}

\begin{thebibliography}{10}
\providecommand{\url}[1]{#1}
\csname url@samestyle\endcsname
\providecommand{\newblock}{\relax}
\providecommand{\bibinfo}[2]{#2}
\providecommand{\BIBentrySTDinterwordspacing}{\spaceskip=0pt\relax}
\providecommand{\BIBentryALTinterwordstretchfactor}{4}
\providecommand{\BIBentryALTinterwordspacing}{\spaceskip=\fontdimen2\font plus
\BIBentryALTinterwordstretchfactor\fontdimen3\font minus
  \fontdimen4\font\relax}
\providecommand{\BIBforeignlanguage}[2]{{%
\expandafter\ifx\csname l@#1\endcsname\relax
\typeout{** WARNING: IEEEtran.bst: No hyphenation pattern has been}%
\typeout{** loaded for the language `#1'. Using the pattern for}%
\typeout{** the default language instead.}%
\else
\language=\csname l@#1\endcsname
\fi
#2}}
\providecommand{\BIBdecl}{\relax}
\BIBdecl

\bibitem{Taxonomy_J3016_202104}
{SAE J3016}, ``Taxonomy and definitions for terms related to driving automation
  systems for on-road motor vehicles - ground vehicle standard: On-road
  automated driving (orad) committee,'' United States, April.2021.

\bibitem{ISO.21448}
\BIBentryALTinterwordspacing
{ISO 21448:2022}, ``Road vehicles -- safety of the intended functionality,''
  Switzerland, 2022-06. [Online]. Available:
  \url{https://www.iso.org/standard/77490.html}
\BIBentrySTDinterwordspacing

\bibitem{NTSB.}
\BIBentryALTinterwordspacing
{National Transportation Safety Board}, ``Collision between a car operating
  with automated vehicle control systems and a tractor-semitrailer truck near
  williston, florida, may 7, 2016.'' [Online]. Available:
  \url{https://www.ntsb.gov/investigations/AccidentReports/Reports/HAR1702.pdf}
\BIBentrySTDinterwordspacing

\bibitem{10.1007/978-981-99-3043-2_15}
M.~Patel, R.~Jung, and Y.~Cakir, ``Simulation-based testing of foreseeable
  misuse by the driver applicable for~highly automated driving,'' in
  \emph{Proceedings of Eighth International Congress on Information and
  Communication Technology}, X.-S. Yang, R.~S. Sherratt, N.~Dey, and A.~Joshi,
  Eds.\hskip 1em plus 0.5em minus 0.4em\relax Singapore: {Springer Nature
  Singapore}, 2024, pp. 183--191.

\bibitem{SOTIF.lanecentering.highwayschauffeursystem}
\BIBentryALTinterwordspacing
C.~Becker, J.~C. Brewer, and L.~Yount, \emph{Safety of the Intended
  Functionality of Lane-Centering and Lane-Changing Maneuvers of a Generic
  Level 3 Highway Chauffeur System}, 2020. [Online]. Available:
  \url{https://rosap.ntl.bts.gov/view/dot/53628}
\BIBentrySTDinterwordspacing

\bibitem{inproceedings}
F.~Warg, S.~Ursing, M.~Kaalhus, and R.~Wiik, ``Towards safety analysis of
  interactions between human users and automated driving systems,'' 2020.

\bibitem{safetyfirst}
\BIBentryALTinterwordspacing
M.~{Wood et al.}, \emph{Safety First For Automated Driving}, 2019. [Online].
  Available:
  \url{https://www.aptiv.com/docs/default-source/white-papers/safety-first-for-automated-driving-aptiv-white-paper.pdf}
\BIBentrySTDinterwordspacing

\bibitem{time.takeover}
A.~Eriksson and N.~A. Stanton, ``Takeover time in highly automated vehicles:
  Noncritical transitions to and from manual control,'' \emph{Human Factors},
  vol.~59, no.~4, pp. 689--705, 2017.

\bibitem{Knapp.2009}
A.~Knapp, M.~Neumann, M.~Brockmann, R.~Walz, and T.~Winkle, \emph{ADAS Code of
  Practice (Code of Practice for the Design and Evaluation of ADAS)}, 2009.

\bibitem{MKRTCHYAN201693}
\BIBentryALTinterwordspacing
{L. Mkrtchyan}, {L. Podofillini}, and {V.N. Dang}, ``Methods for building
  conditional probability tables of bayesian belief networks from limited
  judgment: An evaluation for human reliability application,''
  \emph{Reliability Engineering {\&} System Safety}, vol. 151, pp. 93--112,
  2016. [Online]. Available:
  \url{https://www.sciencedirect.com/science/article/pii/S0951832016000132}
\BIBentrySTDinterwordspacing

\bibitem{Ting.2010}
K.~M. Ting, ``Confusion matrix,'' in \emph{Encyclopedia of Machine Learning},
  C.~Sammut and G.~I. Webb, Eds.\hskip 1em plus 0.5em minus 0.4em\relax Boston,
  MA: {Springer US}, 2010, p. 209.

\end{thebibliography}

\end{document}